\documentclass[prb,twocolumn,showpacs]{revtex4}

\usepackage{amsmath}
\usepackage{graphicx}

\begin{document}

%%%%%%%%%%%%
\title{On the Ordering Instability of Weakly-Interacting Electrons
in a Dirty Metal}

\author{Chetan Nayak and Xiao Yang}

\affiliation{Department of Physics and Astronomy,
University of California at Los Angeles, Los Angeles,
CA 90095-1547}

\date{\today }

\begin{abstract}
In a dirty metal, electron-electron interactions
in the spin-triplet channel lead to singular corrections
to a variety of physical quantities. We show that
these singularities herald the emergence of
ferromagnetism. We calculate the effective
action for the magnetic moment of weakly-interacting
electrons in a dirty metal and show that a state with finite
ferromagnetic moment minimizes this effective action.
The saddle-point approximation is exact in
an appropriate large-$N$ limit.
We discuss the physics of the ferromagnetic state
with particular regard to thermal fluctuations
and localization effects.
\end{abstract}

\pacs{71.10.Hf, 71.10.-w, 71.30.+h, 73.23.-b}

\maketitle

%%%%%%%%%%%%%%%%%%%%%%%%%%%

\section{Introduction.}

%%Def'n of Fink. problem

The presence in a solid of randomly-distributed imperfections
causes its electrons to propagate diffusively
rather than ballistically.
As a result, when two electrons interact, they tarry
in each other's vicinity, and become more strongly
correlated \cite{Altshuler79}.
As shown by Finkelstein \cite{Finkelstein83}
and Castellani, {\it et al.} \cite{Castellani84a,Castellani84b},
this effect leads to an instability of
the diffusive Fermi liquid fixed point (see also
refs. \onlinecite{Belitz94,Pruisken97,Chamon99})
which manifests itself through divergent
corrections to a variety of physical quantities.
The instability exists
for arbitrarily weak coupling and for arbitrarily weak disorder.
Its simplest incarnation is in an electron system in
a weak magnetic field ${\bf B}$ which suppresses
quantum interference effects and relegates
them to subleading order in the dimensionless
resistance\cite{resistance-note} $t$ (measured in units of
$h/e^2$) while interactions
make a contribution at leading order. Since
the relevant interaction is
in the spin-triplet particle-hole channel,
the $g$ factor must be tuned to zero if
${\bf B}\neq 0$ so that there is no Zeeman splitting, which
would cut off the divergences. Though this instability
was identitified nearly 20 years ago, there
is, at present, no consensus on its outcome.

The recent discovery of evidence for a metal-insulator transition in
two-dimensional electron gases (2DEGs) in Si-MOSFETs \cite{Kravchenko,Popovic},
GaAs \cite{GaAs} heterostructures,  and Si-Ge \cite{SiGe}
has led to a re-examination of the problem of
interacting electrons in a random potential. The
interesting phenomena seen
in these experiments occur in a regime in which interactions are strong
(large $r_s$) and quantum interference effects
are likely to be important
because the conductivity is on the order of the quantum
limit $\sim {e^2}/h$ and $B=0$. Though a
number of interesting theories
\cite{Phillips98,Si98,Chakravarty99,Altshuler99,DasSarma99,Chamon01}
have been proposed, none have provided
a complete explanation of the experimental data, which
isn't surprising, given that the presumably
simpler problem of weakly-interacting electrons
in a weak random potential with ${\bf B}\neq 0$, $g=0$
remains unsolved. It is this simpler problem which
is the subject of this paper

%%Obvious solution
Though it is unresolved, the instability caused by the triplet
interaction is, in fact, a crime with only one plausible
suspect: ferromagnetism. Since the diffusive Fermi liquid is
unstable, the system must
settle into some other phase at low temperature;
since the instability is in the
spin-triplet channel, it is presumably one in which
the spins are ordered. The
basic interaction in this problem is exchange,
which is ferromagnetic. There is no available mechanism
which generates antiferromagnetic interactions. To see
this, note that the instability occurs when the conductance is large,
while superexchange occurs in or near an insulating state (since it
arises when the kinetic energy is treated as a perturbation).
There would be RKKY interactions between localized
spins, such as magnetic impurities, if we had any, but we don't.
In the absence of any antiferromagnetic interactions,
we rule out spin density-waves (SDWs)\cite{sdw-note},
spin glass order, and random singlet phases\cite{Bhatt82}.
The latter two furthermore
require large effective disorder while the
instability of interest here is manifest for
small effective disorder\cite{spin-glass-caveat}.

It has been suggested that the instability
of the triplet interaction
signals the formation of `local moments'\cite{Castellani84b}.
In the weak-interaction, weak-disorder limit, however,
the instability of the diffusive Fermi liquid only develops at
large length scales, so the resulting magnetic moments are
anything but `local'. Local moments might be expected when
the energy cost for flipping a spin at short length
scales is very small. This might occur near the
Stoner transition of the clean system \cite{Narozhny00},
where the critical susceptibility, $\chi\sim 1/T^\gamma$,
is of Curie form for a mean-field exponent $\gamma=1$.
There is also a substantial local moment regime with Curie
susceptibility in the large-$U$ Hubbard model 
(because of the weakness of superexchange for large $U$ )
and doped semiconductors
(because of competing ferromagnetic and
antiferromagnetic interactions).
However, these scenarios are not relevant
to the presumably simpler, though still non-trivial,
problem of the weakly-interacting, weakly-disordered
limit.

%%Previous Work

This leaves us with one remaining possibility and,
indeed, a number of authors have proposed that the singularity
of the triplet interaction channel leads to ferromagnetism.
The basic underlying physics is simply that diffusing
electrons interact more strongly because they move
more slowly and, as a result,  the interaction energy gained by (at least partially)
polarizing the system {\it always} outweighs the
single-particle energy cost,
in contrast to a clean system, where the
interaction energy gain is a constant $U$ which must exceed the
single-particle energy cost $1/{N_F}$, i.e. the Stoner criterion
($N_F$ is the density of states at the Fermi energy).
Andreev and Kamenev \cite{Andreev98} proposed such a mechanism
for finite-size systems by computing the disorder
enhancement of the interaction matrix element for a state with two
electrons in single-particle states separated by small energies.
Heuristically, one can say that the two wavefunctions are
concentrated in the same minima of the potential.
Chamon and Mucciolo \cite{Chamon00}
found a ferromagnetic saddle-point
solution of the one-loop effective
action of Finkelstein's theory. Their calculation
generalizes Stoner's by including quantum corrections
resulting from interactions between diffusion modes.
Belitz and Kirkpatrick \cite{Belitz96}
advanced the same hypothesis
when they noted that the phase transition which
they expected to be driven by the growth of the
triplet interaction is of the same form as a transition
to ferromagnetism in a disordered system.

%%Our solution

In this paper, we calculate the effective
action for the magetization in a weakly-interacting
electron system in a slightly dirty metal.
The effective action is calculated to lowest order
in the dimensionless resistance $t$.
It exhibits the physics described above and is clearly unstable to the
formation of a ferromagnetic moment. In principle,
the emergence of some other type of order could
intervene (this effect would have to be non-perturbative in $t$)
before ferromagnetism develops,
but the arguments given above rule out this possibility.
Thus, we conclude that in the limit of small $t$,
the singularity of the spin-triplet interaction channel leads to ferromagnetism.
Our results substantiate the conclusions of previous
authors \cite{Andreev98,Chamon00,Belitz96}.

The basic philosophy is simple: since the RG equations
tell us that the interaction in the triplet channel
grows in importance as the energy scale is lowered,
we try to satisfy this term in the action first by allowing
the magnetization, $X$, to develop a non-zero expectation
value. In order to do this, we compute
an effective action $S_{\rm eff}[X]$
for the magnetization $X$ taking Finkelstein's
`non-linear $\sigma$-model' for
interacting electrons in a disordered system
as the point of departure. Since our goal is
to calculate an effective
action $S_{\rm eff}[X]$ to lowest order
in the resistance $t$, Finkelstein's
$\sigma$-model is merely a convenient
way of organizing the pertinent diagrams.
We believe that it is intuitively appealing to
follow Finkelstein and work directly with a
theory which retains only diffusion modes and
drops the superfluous high-energy electronic
degrees of freedom, but this is a matter of convenience,
and it is a option which we can take regardless of whether new
terms are generated at the two-loop or
higher level (i.e. regardless of whether
the non-linear terms in the theory are really
related by symmetry, as they would be
in a $\sigma$-model). Our derivation
emphasizes the similarity
between, on the one hand, the RG equation for the triplet-channel
interaction in a dirty metal
and the consequent development of ferromagnetism
with, on the other hand,
the RG equation for the Cooper-channel interaction in
a clean Fermi liquid and the consequent development of
superconductivity. We believe that this illuminates the
inevitability of our conclusion.

Once ferromagnetic order has developed,
the system becomes essentially one of spinless
electrons whose diffusion is decoupled
from the spin waves of the system. In
this regime, the suppression of the resistivity
which accompanies the growth of the triplet
interaction parameter is gone. Thus, our
results indicate that a metallic state,
if it exists in two dimensions, does
not owe its existence to the growth
of the triplet interaction parameter.

In two dimensions, our results should be
interpreted as follows. As we lower
the temperature, the system develops local
ferromagnetic order as we cross the mean-field
transition temperature $T_c^{MF}$. The spin density is
controlled by classical thermal
fluctuations which prevent order and lead
to a correlation length $\xi \sim e^{({\rm const.})/T}$
at low-temperatures. In the regime $0<T<T_c^{MF}$,
we have a system in which charge diffuses while the spins
are gapped. The localization physics of
this system is essentially the same as that of spinless
electrons, so the system leaves the diffusive
regime and eventually becomes localized. Thus,
our small $t$ analysis cannot be used all
the way to zero temperature. The
trend is towards a ferromagnetic insulator,
but the onset of insulating behavior can thwart
the development of ferromagnetism. Recent experiments
\cite{Prus02} performed near the
metal-insulator transition indicate
an enhancement of the spin susceptibility
which is cut off by the onset of insulating behavior.
The metal-insulator transition
is outside of the regime of validity of our
calculation, but this is, nevertheless,
qualitatively consistent with our picture.

In $2+\epsilon$ dimensions, when the interaction
strength is larger than a threshhold
value of $O(\epsilon)$, there is a ferromagnetic
transition at a finite temperature $T_c$. If the
resistance is smaller than a
critical value ${t_c}=O(\epsilon)$,
then the resistance will remain
finite down to zero temperature, where
the system is in a
metallic ferromagnetic state.

\section{The Model}
%%Starting effective action

We begin with a system of interacting electrons moving in a quenched
random potential $V(x)$ in $2D$. The imaginary-time
action is
\begin{equation}
S= {S_0} + S_{\rm disorder} + S_{\rm int}
\end{equation}
where
\begin{multline}
{S_0}+S_{\rm disorder} =\\
 T{\sum_{n}}\int {d^2}x \;
{\psi_{n,\alpha}^\dagger}(x)\!
\left(i{\epsilon_n} +
\frac{1}{2m}{\nabla^2} + \mu + V(x)\right){\psi_{n,\alpha}}(x)
\end{multline}
and
\begin{multline}
S_{\rm int} =\\
T{\sum_{n,m,l}}\int \biggl[{\Gamma_1}\,
{\psi_{n,\alpha}^\dagger}({\bf k}+{\bf q}){\psi_{m,\alpha}}({\bf k})
{\psi_{l,\beta}^\dagger}({\bf p}-{\bf q}){\psi_{l+m-n,\beta}}({\bf p})\\
\:+\: {\Gamma_2}\,
{\psi_{n,\alpha}^\dagger}({\bf k}+{\bf q}){\psi_{m,\alpha}}({\bf p})
{\psi_{l,\beta}^\dagger}({\bf p}-{\bf q}){\psi_{l+m-n,\beta}}({\bf k})
\biggr]
\end{multline}
For later convenience, we have written the second term
in momentum space and have split
the interaction into nearly forward scattering, in which
electrons with momenta and spins
$({{\bf k}},\alpha)$, $({{\bf p}},\beta)$
are scattered to $({\bf k'},\alpha)$, $({\bf p'},\beta)$,
where ${\bf k'}\approx{\bf k}$, ${\bf p'}\approx{\bf p}$
and exchange scattering, in which ${{\bf k'}} \approx
{{\bf p}}$, ${{\bf p'}} \approx {\bf k}$.
If the microscopic electron-electron interaction is
$V({\bf q})$, then the associated bare couping constants are
${\Gamma_1}=V(0)$, ${\Gamma_2}=
\int (d\theta/2\pi)\:V(2{k_F}\cos\theta/2)$.
As Finkelstein noted, these are subject
to finite Fermi liquid renormalization
in the ballistic regime, even before they
begin to flow in the diffusive regime.
We assume that the system is in a weak
magnetic field, so that we can ignore
Cooper scattering, for which ${{\bf k}_1} \approx -{{\bf k}_2}$,
${{\bf k}_3} \approx -{{\bf k}_4}$.

In order to perform the average over the quenched random potential,
we replicate the theory and then average the replicated functional
integral over the probability distribution
$\overline{V({\bf x})V({\bf x'})}=
\frac{1}{2\pi{N_F}\tau}\,\delta^{(d)}({\bf x}-{\bf x'})$.
Now, we have
\begin{widetext}
\begin{multline}
S^{\rm rep} = T{\sum_{n;a}}\int {d^2}x\;\biggr[
{\psi_{n,\alpha}^\dagger}(x)
\left(i{\epsilon_n} +
\frac{1}{2m}{\nabla^2} + \mu\right){\psi_{n,\alpha}}(x)
+ \frac{1}{4\pi{N_F}\tau}\,\left({\psi_{n,\alpha}^{a\,\dagger}}
{\psi^a_{n,\alpha}}\right)^2 \biggl]\\
T{\sum_{n,m,l;a}}\int \biggl[{\Gamma_1}\,
{\psi_{n,\alpha}^{a\,\dagger}}({\bf k}+{\bf q})
{\psi^a_{m,\alpha}}({\bf k})
{\psi_{l,\beta}^{a\,\dagger}}({\bf p}-{\bf q})
{\psi^a_{l+m-n,\beta}}({\bf p})
+ {\Gamma_2}\,
{\psi_{n,\alpha}^{a\,\dagger}}({\bf k}+{\bf q})
{\psi^a_{m,\alpha}}({\bf p})
{\psi_{l,\beta}^{a\,\dagger}}({\bf p}-{\bf q})
{\psi^a_{l+m-n,\beta}}({\bf k})
\biggr]
\end{multline}
\end{widetext}
where $a=1,2,\ldots,N$ is a replica index; eventually,
we must take the replica limit, $N\rightarrow 0$.
$N_F$ is the single-particle
density-of-states at the Fermi energy.
The resulting effective action now contains four quartic
terms: one disorder term and two interaction terms. Note that the
disorder term is non-local in time since
it only depends on two Matsubara
frequencies, and it couples different replicas.
The interaction terms depend on three Matsubara
frequencies, so they are local in time,
and they do not couple different replicas.

\section{Finkelstein's Effective Field Theory}

We decouple these quartic terms
with three Hubbard-Stratonovich fields, $Q_{nm,\alpha\beta}^{ab}$,
${Y^a_{nm}}~=~{Y^a}(n\!-\!m)$,
${X^a_{nm,\alpha\beta}}~=~{X^a_{\alpha\beta}}(n\!-\!m)$.
It will be useful to think of $X,Y$ both as matrices
with indices $n,m$ and also as functions of a single
frequency $\omega_{n-m}$; they have only a single replica
index, however, because interactions do not mix different replicas.
Note that $X_{\alpha\beta} \,\vec{\bf \sigma}_{\beta\alpha}$
and $Y$ are essentially the spin and charge density.
Integrating out the fermions, we can write
the partition function as
\begin{equation}
Z = \int {\cal D}Q\,{\cal D}X\,{\cal D}Y \;
{e^{-S_{\rm eff}[Q,X,Y]}}
\end{equation}
where $S_{\rm eff}[Q,X,Y]$ is given by
\begin{widetext}
\begin{multline}
{S_{\rm eff}}[Q,X,Y] =
\sum\int\biggl[-\text{tr}\,\ln\left(i{\epsilon_n}+
\frac{1}{2m}{\nabla^2}+\mu + \frac{i}{2\tau}Q 
+ i{\Gamma_1} \, Y
+{\Gamma_2} \, X \right)
+ \frac{\pi N_F}{4\tau}\text{tr}\left({Q^2}\right)\\ 
+ \, \frac{1}{2}\,{\Gamma_1}\,\text{tr}\left({Y^2}\right)\, +\,
\frac{1}{2}\,{\Gamma_2}\,\text{tr}\left({X^2}\right) 
\biggr]
\label{eq:H-S-action}
\end{multline}

In this expression, we use `tr' to mean
the trace over all of the indices which are not
explicitly written. Matsubara indices are treated
as ordinary matrix indices except that their summations
come with factors of $T$.

 The saddle-point equations are
\begin{equation}
\label{eq:S-P-eqn-Q}
\pi{N_F}\hat{Q}
= i \int \frac{{d^2}p}{(2\pi)^2}\,
\frac{1}{i\hat{\epsilon_n}+\frac{1}{2m}{\nabla^2}\, +
\mu + \frac{i}{2\tau}\hat{Q}+ i{\Gamma_1} \, Y
+{\Gamma_2} \, X}
\end{equation}
\begin{equation}
\label{eq:S-P-eqn-X}
X(m)
=  T{\sum_n}
\int \frac{{d^2}p}{(2\pi)^2}\,
\frac{1}{i{\epsilon_n}{\delta_{n+m,n}}+\frac{1}{2m}{\nabla^2}\, +
\mu + \frac{i}{2\tau}\hat{Q}+ i{\Gamma_1} \, \hat{Y}
+{\Gamma_2} \, \hat{X}}
\end{equation}
\begin{equation}
\label{eq:S-P-eqn-Y}
Y(m)
= i T{\sum_n}
\int \frac{{d^2}p}{(2\pi)^2}\,
\frac{1}{i {\epsilon_n} {\delta_{n+m,n}}+\frac{1}{2m}{\nabla^2}\, +
\mu + \frac{i}{2\tau}\hat{Q}_{n+m,m}+ i{\Gamma_1} \, \hat{Y}
+{\Gamma_2} \, \hat{X}}
\end{equation}

\end{widetext}
The momentum integrals on the right-hand-sides of
(\ref{eq:S-P-eqn-X}) and (\ref{eq:S-P-eqn-Y})
are simply $i\pi{N_F} Q_{n+m,n}$, according to 
(\ref{eq:S-P-eqn-Q}). Hence, they have the solution
\begin{eqnarray}
\label{eq:saddle-sol}
Q_{mn,\alpha\beta}^{ab} &=& 
\text{sgn}\left({\epsilon_n}\right)
\;\delta_{mn}\;\delta_{\alpha\beta}\;\delta^{ab}\cr
{X^a_{\alpha\beta}}(m) &=& 0\cr
{Y^a}(m) &=& 0
\end{eqnarray}
Note that the right-hand sides of the second and third equations
of (\ref{eq:S-P-eqn-X}) and (\ref{eq:S-P-eqn-Y}) vanish
for $m\neq 0$ because
the saddle-point $Q$ is diagonal in Matusubara frequency
while they vanish for $m=0$ because of cancellation between positive
and negative frequencies. As we will see later,
for ${\Gamma_2}$ sufficiently large,
$X=0$ becomes unstable and there is an $X\neq 0$
solution of (\ref{eq:S-P-eqn-X}). This is the
Stoner instability. Upon including low-energy fluctuations
of $Q$ -- diffusion modes -- an $X\neq 0$ solution
arises even for ${\Gamma_2}$ small, as we will
see shortly.

The low-energy fluctuations about this
saddle-point are transverse fluctuations
of $Q$, which can be parametrized
by $V_{nm,\alpha\beta}^{ab}$:
\begin{eqnarray}
\label{eq:Q-param}
{Q} = 
\left(\begin{array}{lccr}
\left(1-V{V^\dagger}\right)^{1/2} & V \\
{V^\dagger} &  -\left(1-{V^\dagger}V\right)^{1/2}\\
                      \end{array} \right)
\end{eqnarray}
The diagonal blocks of $Q_{nm}$ correspond to Matsubara
indices $n>0,m>0$ and $n<0,m<0$ while the upper right
block corresponds to $n<0,m>0$; the lower left block,
to $n>0,m<0$.
These fluctuations correspond to diffusion
of charge and spin. The longitudinal fluctuations
of $Q$ as well as fluctuations of $X$, $Y$ are
gapped at the classical level (i.e. tree-level).

In order to derive an effective field theory
for these diffusion modes, we 
shift $Q\rightarrow Q - 2\tau{\Gamma_1}\,Y
+2\tau i {\Gamma_2} \, X$ in
(\ref{eq:H-S-action})
\begin{widetext}
\begin{multline}
{S_{\rm eff}}[Q,X,Y] =
\sum\int\biggl[-\text{tr}\,\ln\left(i{\epsilon_n}+
\frac{1}{2m}{\nabla^2}+\mu + \frac{i}{2\tau}Q\right)
+ \frac{\pi N_F}{4\tau}\text{tr}\left({Q^2}\right)
- \pi{N_F}{\Gamma_1}\text{tr}\left(QY\right)
+ i\pi{N_F} {\Gamma_2}\text{tr}\left(QX\right)\\
+\,\frac{1}{2}\left(1+2\pi\left(\tau T{\sum_m}\right){N_F}{\Gamma_1}\right)
\,{\Gamma_1}\,\text{tr}\left({Y^2}\right) \,+\,
\frac{1}{2}\left(1-2\pi\left(\tau T{\sum_m}\right){N_F}{\Gamma_2}\right)
\,{\Gamma_2}\,\text{tr}\left({X^2}\right) 
\biggr]
\label{eq:H-S-action2}
\end{multline}

\end{widetext}
Formally, the factors of $\tau\left(T{\sum_m}\right)$ in the
final two terms are infinite as a result of the unrestricted
Matsubara sum. However, upon integrating
out the massive longitudinal modes of $Q$, we see
that they are the first terms in the series
$\tau T{\sum_m} - 2\pi\left(\tau T{\sum_m}\right)^2 + \ldots =
\tau T{\sum_m}/\left(1+2\pi\tau T{\sum_m}\right)$,
which is simply $1/2\pi$. Thus, we replace the
factors of $\tau\left(T{\sum_m}\right)$ by $1/2\pi$
in (\ref{eq:H-S-action2}).
Alternatively, we could expand the
$\text{tr}\,\ln(\ldots)$ about the diffusive saddle-point
and keep only terms up to second order
in $X$ and $Y$. We would then obtain the 
same expressions, namely (\ref{eq:H-S-action2}) with the
factors of $\tau\left(T{\sum_m}\right)$ replaced
by $1/2\pi$. The coefficient of $X^2$ is now
$1-{N_F}{\Gamma_2}$, which becomes negative
for ${\Gamma_2}>1/{N_F}$: the Stoner instability.
Finkelstein actually expanded the
$\text{tr}\,\ln(\ldots)$ only to linear
order in $X$ and $Y$. For small $\Gamma_1$,
$\Gamma_2$, there is no difference, but the
Stoner instabilty is missed.

We now expand the $\text{tr}\,\ln(\ldots)$ about
the diffusive saddle-point. Thus, we require that
$Q$ is a constrained field of the form given in (\ref{eq:Q-param}).
\begin{multline}
\label{eq:almost-sigma-model}
{S_{\rm eff}}[Q] =
\pi{N_F}\int d^d x\; 
\biggl\{D\,\text{tr}{\left(\nabla Q\right)^2}
- 4iZ \text{tr}\left(\hat{\epsilon}Q\right)\\
- {\Gamma_1}\text{tr}\left(QY\right)
+ i {\Gamma_2}\text{tr}\left(QX\right)\\
+\,\frac{1}{2}\left(1+{N_F}{\Gamma_1}\right)
\,\frac{\Gamma_1}{\pi{N_F}}\,\text{tr}\left({Y^2}\right)\\
+\,
\frac{1}{2}\left(1-{N_F}{\Gamma_2}\right)
\,\frac{\Gamma_2}{\pi{N_F}}\,\text{tr}\left({X^2}\right)
\biggr\}
\end{multline}
The diffusion constant $D$ is given by
$D={{v_F^2}{\tau}}/2$.

Following Finkelstein, we integrate out $X,Y$
and obtain an effective action for the diffusion
modes. 
\begin{multline}
\label{eq:sigma-model}
{S_{\rm eff}}[Q] =
\pi{N_F}\int d^d x\; 
\biggl\{D\,\text{tr}{\left(\nabla Q\right)^2}
- 4i \text{tr}\left(\hat{\epsilon}Q\right)\\
-\pi{N_F}{\tilde \Gamma_1}
Q^{aa}_{{n_1}{n_2},\alpha\alpha}
Q^{aa}_{{n_3}{n_4},\beta\beta}
\delta_{n_1-n_2+n_3-n_4}\\
+\pi{N_F}{\tilde \Gamma_2}
Q^{aa}_{{n_1}{n_2}\alpha\beta} 
Q^{aa}_{{n_3}{n_4}\beta\alpha} 
\delta_{n_1+n_2-n_3-n_4}
\biggl\}
\end{multline}
Note that ordinary matrix multiplication rules
reflect the non-locality in time of the first term (the `disorder term') in
this action. The ${\tilde \Gamma_{1,2}}$ interaction
terms do not involve matrix multiplication
and are, consequently, local in time.
In this expression,
${\tilde \Gamma_{1,2}}~=~{\Gamma_{1,2}}/(1~\pm~{N_F}{\Gamma_{1,2}})$.
These corrections to ${\Gamma_{1,2}}$ follow from our retention
of the $X^2$ and $Y^2$ terms which Finkelstein drops. For
${\Gamma_{1,2}}$ small, ${\tilde \Gamma_{1,2}}={\Gamma_{1,2}}$,
and Finkelstein's effective action is recovered.

Initially, the coefficient of the
$\text{tr}\left(\hat{\epsilon}Q\right)$ term
is  $1$, as in (\ref{eq:almost-sigma-model}),
but quantum corrections cause it to flow, so
we have followed Finkelstein in introducing
the coupling $Z$.

\section{Renormalization Group Flows}
%%RG equations

The one-loop RG equations associated with this action are
\begin{eqnarray}
\label{eq:rgeqns}
\frac{dt}{d\ell} &=& {t^2}\left(4 - 3\,\frac{Z_2}{\Gamma_2}
\ln\left(\frac{Z_2}{Z}\right) + \frac{Z_1}{2{\Gamma_1}-{\Gamma_2}}
\ln\left(\frac{Z_1}{Z}\right)\right)\cr
\frac{dZ}{d\ell} &=& t\left(-{\Gamma_1} + 2{\Gamma_2}\right)\cr
\frac{d{\Gamma_1}}{d\ell} &=& t\left({\Gamma_2} +
\frac{\Gamma_2^2}{Z}\right)\cr
\frac{d{\Gamma_2}}{d\ell} &=& t\left({\Gamma_1} +
\frac{2\Gamma_2^2}{Z}\right)
\end{eqnarray}
where $t$ is the dimensionless resistance, $t=1/(2\pi D)$;
${Z_1}=Z-2{\Gamma_1}+{\Gamma_2}$; ${Z_2}=Z+{\Gamma_2}$;
and we have set $\pi{N_F}=1$ in order to avoid clutter.

For sufficiently large interaction strength, the correct
starting point is presumably a Wigner crystal, which
corresponds to a radically different saddle-point
of the effective action (\ref{eq:H-S-action}). Thus, even though
these RG equations were computed to all orders in $\Gamma_{1,2}$
for $t$ small, they should be understood as holding only
for the vicinity of the diffusive saddle-point.

The RG equations for $Z$, ${\Gamma_1}$, ${\Gamma_2}$
can be combined to give:
\begin{equation}
\frac{d}{d\ell}\left(Z-2{\Gamma_1}+{\Gamma_2}\right) = 0
\label{eq:rgward}
\end{equation}
Hence, $Z-2{\Gamma_1}+{\Gamma_2}$ is a constant.
For long-ranged Coulomb interactions, the Ward identity
implies that this constant is zero:
$Z-2{\Gamma_1}+{\Gamma_2}=0$. 
More generally, we can introduce ${\gamma_1}={\Gamma_1}/Z$,
${\gamma_2}={\Gamma_2}/Z$, so that
${\gamma_1}=\frac{1}{2}(1+{\gamma_2}) - {\rm const.}/Z$.
Since $Z$ grows under the RG flow (\ref{eq:rgeqns}),
the final term can be dropped and we have
${\gamma_1}=(1+{\gamma_2})/2$. In the case of spinless
electrons, $Z$ does not grow under the corresponding RG
flow, so we cannot make this simplification; we will
later discuss the similar case of spin-polarized electrons.

Making the above substitution, we find that
the RG equations for $\gamma_2$ and $Z$ are:
\begin{eqnarray}
\frac{d{\gamma_2}}{d\ell} &=& t{\left(1+{\gamma_2}\right)^2}\cr
\frac{dZ}{d\ell} &=& Z\,t\left(-\frac{1}{2}
+\frac{3}{2}\,{\gamma_2}\right)
\end{eqnarray}
Thus, $\gamma_2$ and $Z$ grow as the scale is increased.
Eventually, this growth causes these equations to break down.
If we ignore the flow of $t$, then the breakdown
scale is $E \sim (1/\tau){e^{-\frac{1}{t(1+\gamma_2)}}}$;
in fact, the scale is somewhat lower because $t$ flows.

As noted in ref. \onlinecite{Chamon99}, the breakdown of
these equations is reminiscent of the situation
in a clean Fermi liquid, where the Cooper channel
interaction flows according to $dV/d\ell = -{V^2}$.
The breakdown of the latter at a scale
$\Delta\sim\Lambda {e^{-1/V}}$ signals the onset
of superconductivity or, at least, pairing.
Unlike in the case of a clean Fermi liquid
with a four-fermion Cooper channel interaction,
$Z$ diverges as well in our problem. However, the divergence of
$Z$ is a by-product of the divergence of $\gamma_2$;
In the Cooper pairing problem, if the interaction
which drives pairing could also contribute to
mass renormalization, we would have the same situation
as in the Finkelstein problem. Indeed, this is what happens
when pairing is driven by a gauge field. The development
of pairing not only cuts off the flow of the
interaction, but also the flow of the effective mass
\cite{Bonesteel96}.
Similarly, we expect that the physics which resolves
the divergence of $\gamma_2$ will also resolve
the divergence of $Z$.

\section{Effective Action for Magnetic Degrees
of Freedom: One-Loop}
\label{sec:one-loop-S}

To see how the breakdown of (\ref{eq:rgeqns}) leads to
ferromagnetism, let us go back to the action
(\ref{eq:almost-sigma-model}). Instead of integrating out $X$ and $Y$
at the paramagnetic diffusive saddle point, let's integrate
out $V$ and $Y$ in order to obtain an effective
action $S_{\rm eff}[X]$ for the magnetization $X$.

The spin polarization $X$ acts as a Zeeman
field. A constant field aligned along the $z$ direction,
$X^a_{\alpha\beta} = X {\sigma^z_{\alpha\beta}}$ (taken to be
replica-diagonal), gives a gap to $V_{\uparrow\downarrow}$
and $V_{\downarrow\uparrow}$ but leaves $V_{\uparrow\uparrow}$
and $V_{\downarrow\downarrow}$ unaffected.
To see this, note that a constant field couples only to the diagonal
Matusbara indices $Q_{nn}$. Expanding $(1-V{V^\dagger})^{1/2}$
and $-(1-{V^\dagger}V)^{1/2}$ and retaining only the leading
terms, which are quadratic, we find that 
\begin{equation}
i{\Gamma_2}\,{\rm tr}(XQ) =
i{\Gamma_2}\,X\, {\rm tr}\left[\left(V_{\downarrow\uparrow}
V^\dagger_{\uparrow\downarrow} - V_{\uparrow\downarrow}
V^\dagger_{\downarrow\uparrow}\right)\right]
\end{equation}
An imaginary gap $\pm i{\Gamma_2}X$ in Matsubara formalism corresponds
to a real gap in real-time. $V_{\uparrow\downarrow}$
and $V_{\downarrow\uparrow}$  have propagators:
\begin{multline}
\left\langle V_{nm;\,\uparrow\downarrow,\downarrow\uparrow}({\bf q})
V^\dagger_{nm;\,\downarrow\uparrow,\uparrow\downarrow}(-{\bf q})
\right\rangle =\\
\frac{1}{Z|\omega_{nm}|+D{q^2} \pm i{\Gamma_2}X}
\end{multline}
When we integrate out $V_{\uparrow\downarrow}$ and
$V_{\downarrow\uparrow}$, infrared divergences will be
cutoff by $X$, as the single-particle gap cuts off divergences
in BCS theory. Since $V_{\uparrow\uparrow}$
and $V_{\downarrow\downarrow}$ are unaffected by $X$
to this order, they simply contribute an unimportant constant
to $S_{\rm eff}[X]$.

%%Simplest case (Gamma_1=0, tree level)

We will integrate out $V$
to one loop to obtain $S_{\rm eff}[X]$.
This is not quite the same as a calculation
to lowest order in $t$, but the basic physics is
illustrated most simply in a one-loop calculation,
so we will present it first. To maintain further simplicity,
let us consider the case ${\Gamma_1}=0$; the general
case is a straightforward but more complicated extension,
to which we return later. In the next subsection,
we will present a more careful calculation which
includes the effects of ${\Gamma_1}$ and includes
all of the terms which contribute to lowest order
in $t$.

At one loop, we only need to retain terms in
(\ref{eq:almost-sigma-model}) up to quadratic order
in $V$
\begin{widetext}
\begin{multline}
\label{eq:V-X-action1}
S[V,X] = \pi{N_F}\sum \int \biggl\{
D\, {\rm tr}\left(\nabla{V_+^\dagger} \, \nabla {V^{}_+}\right)
+ Z |{\omega_{nm}}| {V_{nm,+}^\dagger} V_{mn,+}^{}
+ \frac{1}{2} \left(1-{N_F}{\Gamma_2}\right)
\frac{\Gamma_2}{\pi{N_F}}\,{\rm tr}\left({X^2}\right)
\\
+ i{\Gamma_2}\left[ {\rm tr}\left(V^{}_{nm,+} X(n-m)\right)
+ {\rm c.c.}\right]
+ i {\Gamma_2} \left[ {\rm tr}
\left(V^{}_{ml,+} V^\dagger_{ln,+} X(n-m)\right)
+ {\rm c.c.}\right]
+ \left({V_+}\rightarrow{V_-}\, , \: X\rightarrow -X\right)
\biggr\}
\end{multline}
Here, we have written ${V^{}_+}\equiv{V_{\uparrow\downarrow}}$,
${V^\dagger_+}\equiv{V^\dagger_{\downarrow\uparrow}}$,
${V^{}_-}\equiv{V_{\downarrow\uparrow}}$,
and ${V^\dagger_-}\equiv{V^\dagger_{\uparrow\downarrow}}$.

We now shift $X$ in order to remove the bi-linear
coupling between $X$ and $V$. This is the simplest
way to proceed with a one-loop computation of
$S_{\rm eff}[X]$. To go beyond one-loop,
it is more convenient to shift $V$, as we will see
later. Performing the shift $X(l)\rightarrow
X(l)-iT{\sum_{m,A}}{V_{m+l,m;\,A}}$
for $l>0$ and $X(l)\rightarrow X(l)-iT{\sum_{n,A}}
{V^\dagger_{n+l,n;\,A}}$ for $l<0$, where $A=\pm$,
we have:
\begin{multline}
S[V,X] = \pi{N_F}\sum \int \biggl\{
D\, {\rm tr}\left(\nabla{V^\dagger} \, \nabla V\right)
+ Z |{\omega_{nm}}| {V_{nm;\,A}^\dagger} V^{}_{mn;\,A}
- {\Gamma_2} V^\dagger_{nm;\,A} V^{}_{m'n';\,A}\,\delta_{n-m,n'-m'}\\
+ i {\Gamma_2} \left[ {\rm tr}
\left(V^{}_{ml;\,A} V^\dagger_{ln;\,B} {\sigma^z_{AB}}\,X(n-m)\right)
+ {\rm c.c.}\right]
+ \frac{1}{2} \left(1-{N_F}{\Gamma_2}\right)\,
\frac{\Gamma_2}{\pi{N_F}}\,{\rm tr}\left({X^2}\right)
\biggr\}
\label{eq:imp-terms}
\end{multline}
\end{widetext}
A $V^3$ term is also generated as a result of
this shift, but it doesn't contribute to 
$S_{\rm eff}[X]$ at one loop, so we drop it.
The $V$ propagator which follows from (\ref{eq:imp-terms})
is now
\begin{multline}
\left\langle V^{}_{nm;\,\pm;\,ab}({\bf q})
V^\dagger_{nm;\,\pm;\,ba}(-{\bf q})
\right\rangle =
{D^{X\pm}_0}\left(\omega_{nm},q\right)\:+\\
{\Gamma_2}\,{D^{X\pm}_0}\left(\omega_{nm},q\right)\,
{D^{X\pm}_2}\left(\omega_{nm},q\right)\,
\delta_{ab}
\end{multline}
where
\begin{eqnarray}
{D^{X\pm}_0}\left(\omega_{nm},q\right)&=&
\frac{1}{Z|\omega_{nm}|+D{q^2} \pm i {\Gamma_2}X}\cr
{D^{X\pm}_2}\left(\omega_{nm},q\right)&=&
\frac{1}{{Z_2}|\omega_{nm}|+D{q^2} \pm i
{\Gamma_2}X}
\end{eqnarray}
and $Z_2$ is as defined after (\ref{eq:rgeqns}).

\begin{figure}[tbh!]
\includegraphics[width=3.25in]{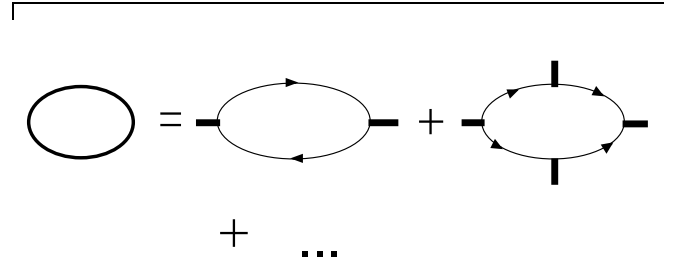}
\caption{The one-loop diagram with $X$-treated as a classical
background field can be viewed as a sum
over one-loop diagrams with $X$ insertions
(the thick lines on the right).}
\label{fig:one-loop-X}
\end{figure}

Integrating out $V$, we obtain all
one-loop diagrams of (\ref{eq:imp-terms}),
with $X$ treated as a classical background field
or, simply, a ${\rm tr}\, \ln(\ldots)$ expression
containing the $V$ propagators in the $X$ background
field. Later, we will return to the possibility of other
saddle-points, but for now, let us assume,
in the name of simplicity, that $X$ is static,
translationally-invariant, and replica-symmetric.
For such $X$, the effective action takes the form:
\begin{multline}
\label{eq:S-eff-X-1}
S_{\rm eff}[X] = {\sum}\int 
\frac{1}{2}\left(1-{N_F}{\Gamma_2}\right)\,
{\Gamma_2}\,
{\rm tr}\left({X^2}\right)\\
+ {\rm tr}\ln\biggl(\left[Z|\omega_{nm}|+D{q^2}\pm i
{\Gamma_2}{X}\right]
\delta_{nn'}\,\delta_{mm'}\,\delta_{aa'}\,\delta_{bb'}\\
- \: {\Gamma_2}\: \delta_{n-m-n'+m'}\,
\delta_{aa'}\,\delta_{bb'}\,\delta_{ab}\biggr)
\end{multline}
This may be rewritten in terms of the propagators
${D^{X\pm}_{0,2}}$ as
\begin{widetext}
\begin{multline}
\label{eq:S-eff-X-2}
S_{\rm eff}[X] = {\sum}\int 
\frac{1}{2}\left(1-{N_F}{\Gamma_2}\right)\,
{\Gamma_2}\,
{\rm tr}\left({X^2}\right)
+ {\rm tr}\ln\biggl[
\left({D^{X+}_0}\:-\:
{\Gamma_2}\,{D^{X+}_0}\,{D^{X+}_2}\,\delta_{ab}\right)
\left({D^{X-}_0}\:-\:
{\Gamma_2}\,{D^{X-}_0}\,{D^{X-}_2}\,\delta_{ab}\right)
\biggr]
\end{multline}
\end{widetext}
When we take the product of the two factors on
the right-hand side of (\ref{eq:S-eff-X-2}),
the ${D^{X+}_0}{D^{X-}_0}$ term
has a free summation over the replica index
$b$ which vanishes in the replica limit
$N\rightarrow 0$ for replica-symmetric solutions
${X_a}=X$. Dropping this term and using the relation 
${D^{X\pm}_0} - {\Gamma_2} |\omega_{nm}| {D^{X\pm}_0}{D^{X\pm}_2}
= {D^{X\pm}_2}$, we have
\begin{multline}
\label{eq:S-eff-X-3}
S_{\rm eff}[X] = {\sum}\int \biggl\{
\frac{1}{2}\left(1-{N_F}{\Gamma_2}\right)\,
{\Gamma_2}\,
{\rm tr}\left({X^2}\right)\\
- {\rm tr}\ln\left({D^{X+}_2}{D^{X-}_2} - {D^{X+}_0}{D^{X-}_0}
\right)
\biggr\}
\end{multline}
Evaluating the integral, we find
\begin{multline}
\label{eqn:S_eff-X-4}
S_{\rm eff}[X] =  \beta {L^2}\Biggl[
\frac{1}{2}\left(1-{N_F}{\Gamma_2}\right)\,
{\Gamma_2}\,
{X^2}\,-\\
t\,\pi{N_F}{\Gamma_2^2}\,
\left(\frac{1}{Z} - \frac{1}{Z_2}\right)\,{X^2}
\left(\ln\left(\frac{\Lambda}{{\Gamma_2}X}\right)-1\right)
\Biggr]
\end{multline}
where $L^2$ is the area of the system.

As a result of the logarithm on the second line,
the minimum of this effective action is at non-zero
$X$ for arbitrarily small $\Gamma_2$.
We will discuss this non-trivial solution
and the saddle-point equation (or gap equation)
which it satisfies in more detail in the next section.
At this stage, the important lesson is that, at the one-loop level,
the weakly-interacting diffusive electron gas
is always unstable to the development of
non-zero $X$, i.e. ferromagnetism.

\section{Ferromagnetic Saddle-Point at One-Loop}

Let us examine more closely the saddle-point equation
for $X$ in the simplest setting, that of section \ref{sec:one-loop-S}.
If we vary equation (\ref{eq:S-eff-X-2}) with respect to $X$
(before performing the momentum integral and Matsubara sum),
then we obtain the saddle-point equation:
\begin{widetext}
\begin{eqnarray}
\label{eq:saddle-point-simple2}
1-{N_F}{\Gamma_2}
&=& \pi{N_F}{\Gamma_2} \sum\int\biggl[
{D^{X+}_2}{D^{X-}_2} - {D^{X+}_0}{D^{X-}_0}\biggr]\cr
&=& \pi{N_F}{\Gamma_2} \sum\int\biggl[
\frac{1}{{\left[{Z_2}|\omega_{nm}|+D{q^2}\right]^2}
+ {\Gamma_2^2}{X^2}} \:-\:
\frac{1}{{\left[Z|\omega_{nm}|+D{q^2}\right]^2}
+ {\Gamma_2^2}{X^2}}
\biggr]\cr
&=& t \,\pi{N_F}{\Gamma_2}\,
\left(\frac{1}{Z} - \frac{1}{Z_2}\right)
\ln\left(\frac{\Lambda}{{\Gamma_2}X}\right)
\end{eqnarray}
\end{widetext}
The integrals on the right-hand-side of the second line
are, for $X=0$, precisely the integrals which determine
the contributions of diagrams (d) and (e) to
the RG equation for $\Gamma_2$. We will further discuss
this point below.

The right-hand-side of the saddle-point equation
is logarithmically divergent in the $X\rightarrow 0$ limit.
Hence, there is always a solution
\begin{equation}
{\Gamma_2}X =
\frac{1}{\tau}\,
e^{-\left(1-{N_F}{\Gamma_2}\right)({1+{\gamma_2}})/({t\,\gamma_2^2})}
\end{equation}
We have taken $\Lambda=1/\tau$ as the ultraviolet cutoff
and ${\gamma_2}=\pi{N_F}{\Gamma_2}/Z$. This is the non-trivial
minimum of the action to which we referred at the end of
section \ref{sec:one-loop-S}. The structure of this
saddle-point equation is clearly reminiscent of
the BCS gap equation. The integrals on the right-hand-side
are logarithmically divergent -- as expected for a marginally
relevant interaction. A symmetry-breaking order parameter must
develop in order to cut them off.

\begin{figure}[tbh!]
\includegraphics[width=3.25in]{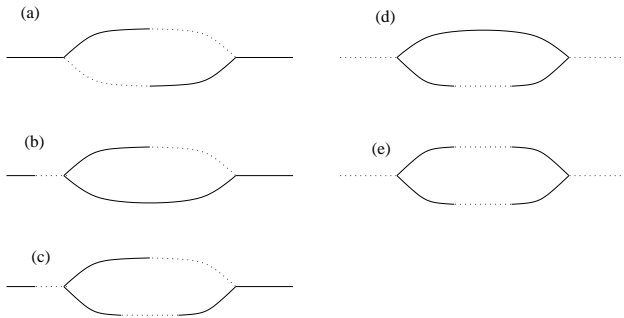}
\caption{The five basic diagrams which contribute
to the renormalization of $\Gamma_2$. Solid lines
are $V$ propagators. Dotted lines are
bare $X$ propagators.}
\label{fig:five-diagrams}
\end{figure}

We can submit our saddle-point solution to an important
check by verifying its consistency with the RG
equations. The gap to spin-flip excitations, ${\Gamma_2}X$
is an observable quantity, and should be independent of the ultraviolet
cutoff, $\Lambda$. Even though the cutoff is taken to
be a physical scale $1/\tau$, we are free to
integrate out short-wavelength fluctuations
in order to lower the cutoff while simultaneously
renormalizing the couplings $\Gamma_1$, $\Gamma_2$,
$t$, and $Z$. This must occur in such a way
that the saddle-point solution, ${\Gamma_2}X$, remains unchanged.
For this reason, we regard the one-loop calculation
of section \ref{sec:one-loop-S} as incomplete. It
is consistent with an RG equation which only
includes diagrams (d) and (e) of figure \ref{fig:five-diagrams}
for the flow of $\Gamma_2$. Diagrams (a), (b), and (c)
are not incorporated, and neither is the flow
of $\Gamma_1$, $t$, $Z$.

\section{Beyond One-Loop}
\label{sec:order-t-S}

Though the preceeding one-loop calculation captures the basic physics
of the problem, it does not capture all of the order $t$
physics. As we have just noted, the effective action (\ref{eq:S-eff-X-1})
only includes terms corresponding to two of the five
diagrams in figure \ref{fig:five-diagrams} -- namely, (d)
and (e) -- which contribute to (\ref{eq:rgeqns}).

The apparent contradiction stems from the fact that
the physics of a one-loop RG equation might only appear
at higher-loops in a saddle-point calculation. This is surprising
at first glance, but can be understood at a technical level
by noting that a contribution to the effective action of
the form ${t^2}{\ln^2}(\Lambda/X)$ is equally important
at the saddle-point as a contribution of the form 
${t}{\ln}(\Lambda/X)$ since the extra logarithm
compensates for the extra factor of $t$.
A simple example of a model in which this occurs
is given in the appendix.

\begin{figure}[tbh!]
\includegraphics[width=3.25in]{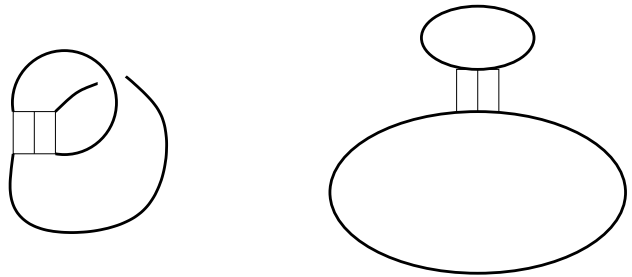}
\caption{Two-loop contributions to $S_{\rm eff}[X]$
which renormalize $\Gamma_2$, $t$, and $z$.
The first diagram renormalizes $\Gamma_2$
while the second renormalizes $t$ and $z$.
The quadratic vertices are $\Gamma_1$ interaction
vertices.}
\label{fig:Gamma_1-renorm}
\end{figure}

The effects of $\Gamma_1$ on the flow of $\Gamma_2$, $t$, $Z$
enter in a two-loop calculation of the effective action.
Consider the diagrams of figure \ref{fig:Gamma_1-renorm}.
These are two-loop diagrams which, when combined with
the one-loop diagram of the previous section,
have the effect of shifting in the effective action of
the previous section the values of $\Gamma_2$, $t$, and $Z$
from their bare values. These diagrams can receive contributions
from non-spin-flip diffusion propagators which
are not cut off by the magnetization,
$X$. They are charge and $S_z$ diffusion modes. Since we expect
them to be a part of the low-energy theory, we can choose not to integrate
them out or to integrate out modes above a cutoff of our choosing.
The natural choice is to integrate out these modes above
${\Gamma_2}X$ so that the resulting effective field theory
for $X$, $V_{\uparrow\uparrow}$, $V_{\downarrow\downarrow}$
contains only excitations with energies less than ${\Gamma_2}X$.
(We could alternatively imagine performing the calculation at
finite-temperature so that $T$ is effectively the cutoff for both
spin-flip and non-spin-flip propagators. Such a calculation could
be used to obtain the transition temperature $T_c$ for the putative
transition.) This effects the following replacements:
\begin{eqnarray}
\label{eqn:two-loop-contribution}
{\Gamma_2} &\rightarrow& {\Gamma_2} +
t{\Gamma_1}\,\ln\!\left(\frac{\Lambda}{{\Gamma_2}X}\right)\cr
t &\rightarrow& t + {t^2} {\Gamma_1}\,{f_2}(z,{z_1},{z_2})\,
\ln\!\left(\frac{\Lambda}{{\Gamma_2}X}\right)\cr
Z &\rightarrow& Z - t{\Gamma_1}\,\ln\!\left(\frac{\Lambda}{{\Gamma_2}X}\right)
\end{eqnarray}
where
\begin{eqnarray}
{f_2}(z,{z_1},{z_2}) &=& \frac{2{z_1}}{{z_1}-{z_2}}\,{f_1}(z,{z_1})
- \frac{2{z_2}}{{z_1}-{z_2}}\,{f_1}(z,{z_2})\cr
{f_1}(z,{z_2}) &=& \frac{1}{{z_2}-{z}}\,\ln\!\left(\frac{z_2}{z}\right)
\end{eqnarray}

Diagrams (a)-(c) of figure \ref{fig:five-diagrams}
cannot contribute directly to $S_{\rm eff}[X]$
for static $X$ because the external $V$ lines cannot carry zero
frequency. The Matsubara indices $m$, $n$ of
$V_{mn}$ necessarily satisfy $n>0$, $m<0$, so their difference
cannot vanish. For the same reason,
these diagrams cannot be linked with (d)
and (e), in the way demonstrated in the toy model
in the appendix. It is possible that contributions
related to diagrams (a)-(c) enter $S_{\rm eff}[X]$ through
even higher-loop contributions or, perhaps, when fluctuations
in $X$ are included. However, this touches upon the
unresolved issue of the renormalizability of the theory at higher-loops.
In the absence of a clearer understanding of such higher-loop
contributions to $S_{\rm eff}[X]$, we conclude that
the one-loop effective action (\ref{eqn:S_eff-X-4}) with the replacement
(\ref{eqn:two-loop-contribution}) is the most systematic
result which we can obtain at present. We expect that
going beyond one-loop in $V$ or including the effects of
fluctuations of $X$ will quantitatively but not qualitatively
modify our result.

In section \ref{sec:one-loop-S}, we removed the bi-linear coupling
between $X$ and $V$ by shifting $X$.
It may seem more natural to remove the bi-linear coupling between
$X$ and $V$ by performing the shift $V_{nm}\rightarrow V_{nm} -
{D_0^{-1}}(q,{\omega_{nm}})X(n-m)$. If we then integrate out
$V$, we will not find a non-trivial replica-symmetric
saddle-point (though there may be replica-symmetry-breaking
saddle-points), and the resulting action will not contain
any of the physics of the one-loop RG equations.
In order to recover this physics, we must now go
to one-loop in $S_{\rm eff}[X]$. Going to one-loop
in $S_{\rm eff}[X]$, we recover the saddle-point of
our initial action of section \ref{sec:one-loop-S}.
The relevant diagrams are diagrams (d) and (e) of
figure \ref{fig:five-diagrams}, but the solid lines
are shrunk to a point when these diagrams are
viewed as $1$-loop diagrams in $S_{\rm eff}[X]$.
Diagrams corresponding to (a), (b), and (c) have
dotted lines attached to the solid external legs
(via the bi-linear coupling between
$X$ and $V$); however, these vanish in the static limit
for the same reason as above. Thus,
at this level, there is no advantage in proceeding this way,
and there is a definite disadvantage in that one
extra step is necessary. However, it is possible
that a higher-loop calculation is simpler
in this approach.

The one-loop effective action which we computed
in section \ref{sec:one-loop-S} is actually
exact in a generalization of the model.
The most naive large-$N$ version of
the model is not particularly simple. If
we simply introduce $N$ flavors of electrons
$\psi_\alpha$, $\alpha=1,\ldots,N$, then the $Q$-matrix has two
flavor indices, $Q_{AB}$, and the large-$N$
limit allows us to drop all non-planar diagrams.
However, we do not know how to sum all of
the planar diagrams. Instead, we consider the
following generalization: we suppose that there
are $N$ flavors of diffuson fields $V_\alpha$, 
$\alpha=1,\ldots,N$. This generalization does not
have a simple interpretation in terms of the original
electron variables, but it is technically simple.
The action (\ref{eq:imp-terms}) is generalized to:
\begin{widetext}
\begin{multline}
S[V,X] = \pi{N_F}\sum \int \biggl\{
D\, {\rm tr}\left(\nabla{V^\dagger} \, \nabla V\right)
+ Z |{\omega_{nm}}| {V_{nm;\,A\alpha}^\dagger} V^{}_{mn;\,A\alpha}
- {\Gamma_2} V^\dagger_{nm;\,A\alpha} V^{}_{m'n';\,A\alpha}\,
\delta_{n-m,n'-m'}\\
+ i {\Gamma_2} \left[ {\rm tr}
\left(V^{}_{ml;\,A\alpha} V^\dagger_{ln;\,B\alpha}
{\sigma^z_{AB}}\,X(n-m)\right)
+ {\rm c.c.}\right]
+ \frac{N}{2} \left(1-{N_F}{\Gamma_2}\right)\,
\frac{\Gamma_2}{\pi{N_F}}\,{\rm tr}\left({X^2}\right)
- \frac{1}{N}\,{\Gamma_2} {\sum_\beta}\left[ {\rm tr}
\left(V^{}_{\alpha} V^\dagger_{\alpha} V^{}_{\beta}\right)
+ {\rm c.c.}\right]
\biggr\}
\label{eq:imp-terms-large-N}
\end{multline}
\end{widetext}
Note that there is still only a single $X$ field. The explicit factor of
$N$ in front of the penultimate term and the factor of
$1/N$ in front of the final term ensures that every term in
the action is $O(N)$. For this action,
the diagram which we calculated in section \ref{sec:one-loop-S}
is the sole $O(N)$ diagram. All others are $O(1)$ or smaller.
Furthermore, our effective action is multiplied by a coefficient
$N$. Thus, in the $N\rightarrow\infty$ limit, our
one-loop saddle-point calculation is exact.

At first glance, it may seem almost contradictory to
take an $N\rightarrow\infty$ limit while
simultaneously taking a replica limit
$n\rightarrow 0$. However, the flavor index structure
in the action (\ref{eq:imp-terms-large-N}) is very
different from the replica index structure. The latter
is chosen to ensure that closed fermion loops
vanish while the latter selects a preferred set of the
remaining diagrams. Hence, the two limits are consistent.

\section{Physics of the Ferromagnetic Diffusive Fermi Liquid}

Having concluded that the divergence of the triplet
interaction amplitude leads to ferromagnetism,
the obvious next question is ``so then what?''
To answer this, we must examine the diffusion of
electrons in a ferromagnetic metal, and the interaction
between diffusion modes and spin waves.

In the ferromagnetic state, we can expand
$X$ about its perfectly-ordered ground state,
which we take to be along the $\hat{\bf z}$-direction
\begin{equation}
{\bf X} = \left({m_x},{m_y},\sqrt{1-{m_x^2}-{m_y}^2}\right)
\end{equation}
We can also perform a gradient expansion
of $S_{\rm eff}[{\bf X}]$. The result is
essentially dictated by symmetry:
\begin{equation}
S_{\rm eff}[{\bf X}] = \int \left( \epsilon_{ij} 
{m_i}{\partial_\tau}{m_j} +
K\left(\nabla{m_i}\right)^2 +
O\left({\left(\nabla{m}\right)^4}\right)\right)
\end{equation}
where $i=1,2$ are the $\hat{\bf x}$ and $\hat{\bf y}$
components of ${\bf X}$.
The constant $K$ is the spin stiffness of the ferromagnet.

We now reinstate the coupling of the order parameter
to the diffusion modes of the system, as in
(\ref{eq:almost-sigma-model}) or (\ref{eq:V-X-action1}).
\begin{multline}
\label{eq:V-m-action1}
S[V,m] = \pi{N_F}\sum \int \biggl\{
D\, {\rm tr}\left(\nabla{V^\dagger} \, \nabla {V^{}}\right)
+ Z |{\omega_{nm}}| {V_{nm}^\dagger} V_{mn}^{}\\
+ i {\Gamma_2}\,{X_0}
\left[{\rm tr}\left({\sigma_z}V^{}_{nm} V^{\dagger}_{mn}\right)
+ {\rm c.c.}\right]\\
+ i {\Gamma_2}\,
\left[{m_i}(\omega_{n-m})\: {\rm tr}\left({\sigma_i}V^{}_{nm} \right)
+ {\rm c.c.}\right]\biggr\}\\
+ {\sum_n}\int \left( \epsilon_{ij} 
i{\omega_n}{m_i}{m_j} +
K\left(\nabla{m_i}\right)^2 \right)
\end{multline}
where $\sigma_i$ are Pauli matrices and we have dropped higher-order
terms in $m_i$.
A purist may object that the $V$'s should not
appear in the action for $m_i$
since the dynamics for the order parameter
was obtained by integrating out the $V$'s.
If we now calculate with this action, we run the risk
of double-counting the degrees of freedom of the system.
In order to avoid this, we must either calculate only to tree-level
in $V$ (but allow loops in $m_i$) or introduce counter-terms
which cancel the final line of (\ref{eq:V-m-action1}). In the
second approach, by computing loops in $V$, we will
cancel the $m_i$ counterterms.

The physics of this model is not completely
manifest in the above form (\ref{eq:V-m-action1})
because a uniform rotation of the ferromagnetic order
parameter appears to be an excitation -- since
there are non-derivative terms in $m_i$ -- when, in fact,
it isn't. This can be made manifest by rotating
the quantization axis of the diffusing electrons so
that it is aligned along the local direction of the
order parameter. This is done by enacting the variable
change
\begin{equation}
V \rightarrow e^{i\epsilon_{kl} {m_k} {\sigma_l}}\, V \,
e^{-i\epsilon_{ij} {m_i} {\sigma_j}}
\end{equation}
The action is now
\begin{multline}
\label{eq:V-m-action2}
S[V,m] = {N_F}\sum \int \biggl\{
D\, {\rm tr}\left(\nabla{V^\dagger} \, \nabla {V^{}}\right)
+ Z |{\omega_{nm}}| {V_{nm}^\dagger} V_{mn}^{}\\
+ i {X_0}{\Gamma_2}\,
\left[{\rm tr}\left({\sigma_z}V^{} V^{\dagger} \right)
+ {\rm c.c.}\right]\\
D\, {\rm tr}\left(\nabla{V_{nm}^\dagger} \:
\nabla\!\left(\epsilon_{ij} {m_i}(\omega_{n-l})
\left[{\sigma_j},{V_{ml}^{}}
\right]\right)\right) + {\rm c.c.}\biggr\}
\\
+ {\sum_n}\int \left( \epsilon_{ij} 
i{\omega_n}{m_i}{m_j} +
K\left(\nabla{m_i}\right)^2 \right)
\end{multline}

As noted earlier, there is a gap for spin-flip
diffusion modes, as given by the second line
of (\ref{eq:V-m-action2}). Thus, we can ignore
these modes at low energies and focus on
charge diffusion and $S_z$ diffusion; $S_{x,y}$
are carried by spin waves $m_{x,y}$. From the structure
of the interaction term on the third
line of (\ref{eq:V-m-action2}), we see that
$m_{x,y}$ cannot couple directly to charge
or $S_z$ diffusion alone, as dictated by
rotational invariance. A charge or $S_z$ diffusion mode,
upon emitting or absorbing a spin wave, must then
become a spin-flip diffusion mode. Thus, this coupling can be
ignored at energies below the gap for spin-flip diffusion modes.
Even in the absence of such a gap, the coupling would
have dimension $d/2$ in $d$-dimensional space
and be strongly irrelevant,
as expected for interactions with Goldstone modes --
which must occur through derivative couplings
such as that on the third
line of (\ref{eq:V-m-action2}).

Thus, the low-energy physics of the ferromagnetic
diffusive metal state is that of ferromagnetic
spin waves decoupled from diffusion in a Zeeman field.
The latter has been well-studied \cite{Castellani84a,Belitz94},
and is similar to the spinless electron problem. Interactions
suppress the conductivity, as may be seen from the
RG equations:
\begin{eqnarray}
\label{eq:rgeqns-polarized}
\frac{dt}{d\ell} &=& {t^2}\Biggl(2 -
\frac{Z-{\Gamma_H}-{\Gamma_1}}{{\Gamma_1} + {\Gamma_H}}
\ln\left(\frac{Z}{Z-{\Gamma_H}-{\Gamma_1}}\right)\cr & &
{\hskip 1 cm} -\,
\frac{Z+{\Gamma_H}+{\Gamma_1}}{{\Gamma_1}-{\Gamma_H}}
\ln\left(\frac{Z-{\Gamma_H}+{\Gamma_1}}{Z}\right)\Biggr)\cr
\frac{dZ}{d\ell} &=& - t {\Gamma_H}\: , \:\:
\frac{d{\Gamma_1}}{d\ell} = 0\: , \:\:
\frac{d{\Gamma_H}}{d\ell} = -t{\Gamma_H}
\end{eqnarray}
${\Gamma_H}$ is the part of the interaction between
${S_z}=0$ diffusion modes which includes a contribution
from both the direct and exchange channels,
${\Gamma_H}\left(Q_{\uparrow\uparrow} Q_{\uparrow\uparrow} +
Q_{\downarrow\downarrow} Q_{\downarrow\downarrow}\right)$;
its bare value is ${\Gamma_H}={\Gamma_1}-{\Gamma_2}$.

In two dimensions, these RG equations describe a flow
towards insulating behavior.
However, there is a metal-insulator transition
in $2+\epsilon$ dimensions through which the
system can be tuned by varying the bare resistivity
through a critical value of $O(\epsilon)$.
We examine the situation
in some more detail in the next section.

The preceeding discussion is strictly true
at zero temperature. At finite temperatures,
thermal fluctuations will destroy
two-dimensional long-range ferromagnetic order,
but some signature of the zero-temperature ferromagnetic
phase persists. If a finite external magnetic field is applied,
the magnetization is a scaling function of temperature and magnetic
field. A large-N calculation for an $SU(N)$
magnet theory gives the following expression for
the scaling function for the susceptibility\cite{Read95}:
\begin{equation}
\chi(r,h) = \frac{M_{0}}{8\pi r}\,
\ln\biggl[(q-e^{-h/2})/(q-e^{h/2}) \biggr]
\label{eqn:suscept-scaling}
\end{equation}
where $r=K/T$ and $h=H/T$ are the rescaled spin stiffness and magnetic field
respectively, $q$ is a root of the equation $(q-e^{-h/2})(q-e^{h/2}) = q^{2}
e^{-8\pi r}$, and $M_{0}$ is the zero temperature magnetization. The
susceptibility is given by $M_{0}(e^{4\pi r}-1)/(8\pi Tr)$,
which diverges exponentially as $T\rightarrow 0$.

In the limit of weak interactions and weak disorder, there
is a substantial temperature regime in which there is
ferromagnetic order with a long -- but finite -- correlation
length and the resistivity is small. If
the dimensionless resistivity, $t$, is taken to be $0.3$
at temperature $5$ K, and $r_{s}$, the average interspace
between electrons measured in the unit of effective Bohr
radius, is taken to be 0.1, then the susceptibility
is large below $15$ mK. However, the resistivity
becomes $O(1)$ at temperatures $\ll 10^{-20}\,K$. While
ferromagnetism sets in at very low temperatures,
insulating behavior is completely academic.
In the temperature interval between these two
scales, we expect the susceptibility to
be of the above form (\ref{eqn:suscept-scaling}).
When $r_s$ is larger, the separation between these scales
decreases, but we should also be more circumspect about
using this weak-coupling theory. Of course,
if the magnetic field is removed, we pass from the
unitary ensemble to the orthogonal one, and
weak-localization effects should drive the
system insulating much more rapidly.

\section{Phase Diagram in $2+\epsilon$ Dimensions}

In $2+\epsilon$ dimensions, the renormalization group
equations are modified by the appearance of
tree-level dimensions associated with the spatial
dimensionality:
\begin{eqnarray}
\label{eq:rgeqns-epsilon}
\frac{dt}{d\ell} &=& -\epsilon t + {t^2}\left(4 - 3\,\frac{Z_2}{\Gamma_2}
\ln\left(\frac{Z_2}{Z}\right) + \frac{Z_1}{2{\Gamma_1}-{\Gamma_2}}
\ln\left(\frac{Z_1}{Z}\right)\right)\cr
\frac{dZ}{d\ell} &=& t\left(-{\Gamma_1} + 2{\Gamma_2}\right)\cr
\frac{d{\Gamma_1}}{d\ell} &=&  t\left({\Gamma_2} +
\frac{\Gamma_2^2}{Z}\right)\cr
\frac{d{\Gamma_2}}{d\ell} &=& t\left({\Gamma_1} +
\frac{2\Gamma_2^2}{Z}\right)
\end{eqnarray}
By varying $\Gamma_2$, we can tune the system
through a paramagnet-ferromagnet
transition.

For $t$, $\gamma_2$ small, the flows are into
the fixed line $t=0$. This fixed line
terminates at ${\Gamma_2}={\Gamma_2^{\rm Stoner}}$.
For $\Gamma_2=0$, there is a metal-insulator transition
at $t=\epsilon$. In the vicinity of this
transition, $\gamma_2$ is relevant. If it flows
to sufficiently large values, then ferromagnetism
will develop. However, if $t$ flows to zero
before this happens, then the system will remain
paramagnetic. Both possibilities can occur
as soon as $\gamma_2$ becomes nonzero, so
the ferromagnet-paramagnet boundary
which begins on the $t=0$ axis at
${\Gamma_2}={\Gamma_2^{\rm Stoner}}$
terminates at $t=\epsilon,{\gamma_2}=0$.
Ferromagnetism, if it occurs,
will have a finite onset temperature. At lower temperatures,
the conductivity may be metallic or insulating, depending
on $t$. In the former case, we have a ferromagnetic
metal. In the latter case, the system becomes insulating
at lower temperatures. It is possible that
the ferromagnetic order is destroyed by the onset
of insulating behavior. This cannot be resolved by
the methods in this paper; thus, we have left
the phase boundary of the paramagnetic insulating state
intentionally vague. (In fact, it may not be
paramagnetic, but might instead be a spin glass or some other
non-ferromagnetic phase.)

\begin{figure}[tbh!]
\includegraphics[width=3.25in]{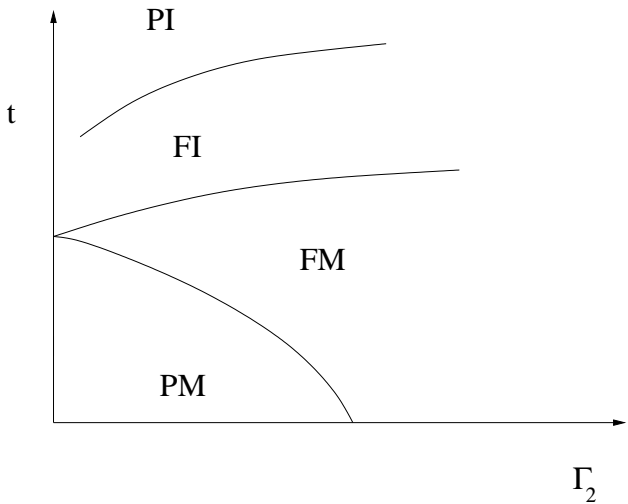}
\caption{Phase diagram in $2+\epsilon$ dimensions. PM is the
paramagnetic metal phase; FM, the ferromagnetic metal phase;
FI, the ferromagnetic insulator. PI is a non-ferromagnetic
insulator: possibly paramagnetic, but perhaps a spin glass.}
\label{fig:pm-fm-fi}
\end{figure}

\section{Discussion}

In this paper, we have used a straightforward
and absolutely standard technique -- albeit
in a somewhat complicated and technically-involved
context -- to argue that
the instability of the weakly-interacting electron
system in a weak random potential signals the development
of ferromagnetism. Since there is a single instability occurring
in the weak-interaction limit and it occurs while
the resistivity is small, we are afforded a great deal
of control over this problem. The instability bespeaks of
the relative importance of one term in the action compared
to the others. We need only satisfy this term --
which we do by allowing for the development of a ferromagnetic moment --
without worrying about other terms,
since none of them are relevant. The demonstrable absence of
a competing instability is the main justification
for a simple mean-field theory. This may be formalized in
a large-$N$ generalization of the problem. Further credence is
given by the physical arguments of the introductory section
which rule out any other resolution of the divergence associated
with the triplet amplitude. This is, in fact, a remarkable
state of affairs, given that there is no comparable
controlled theory for itinerant electron ferromagnetism
in clean metals. In a clean metal, the Stoner instability
occurs at strong coupling, where there are other competing
instabilities and no unambiguous way of deciding between them.

It is striking how little is known about the phase
diagram of the 2DEG. Even at zero magnetic field,
the phase diagram as a function of $r_s$ and disorder strength
is only known in the small and large $r_s$ limits
for vanishing disorder. Intermediate $r_s$ is poorly
understood, as is finite disorder. We believe that
we have, in this paper, extended {\it terra cognita}
to include the small $r_s$, weak disorder region.

We emphasize that our result does not shed too much light
on the metal-insulator transition which is observed in
strongly-interacting 2DEGs. The development of ferromagnetism
does not lead to a metallic state in two dimensions. Once
ferromagnetism has developed, weak localization (which we have
neglected in this paper) and interaction
corrections both drive the system insulating. There is no
contradiction with experiments, however, since the transition
seen there is in the strong-interaction, high-resistivity
($t\sim 1$) regime where our result is not expected to
be valid. On the other hand, our result {\it does} rule
out theories which conjecture
that the divergence of the triplet interaction leads to
a metallic state {\it at zero-temperature} in 2D.

\begin{acknowledgments}
We would like to thank C. Chamon, A. Kamenev,
E. Mucciolo, L. Radzihovsky, R. Shankar, and T. Senthil for discussions.
This work was supported by the National Science Foundation under
Grant No. DMR-9983544 and by the A.P. Sloan Foundation.
\end{acknowledgments}

\appendix

\section{Correspondence Between RG Equations and an Effective Action}

When we transform an action by introducing
Hubbard-Stratonovich fields and integrating out the
original variables of the action, the result can be
an action which looks very different from the original one.
However, the coupling constants of the two actions
are related, and the corresponding RG equations should
be as well. In the simplest cases, this is completely
transparent. Consider, for instance, a system of $N$
flavors of spinless fermions in one-dimension. Suppose that
their density is commensurate with the underlying crystal lattice
and that their interaction respects the $U(N)$ symmetry which
mixes the fermions:
\begin{multline}
\label{eq:GN-model}
{\cal L} = {\psi^\dagger_{aR}}i\left({\partial_t}+{\partial_x}\right)
{\psi^{}_{aR}}
+ {\psi^\dagger_{aL}}i\left({\partial_t}-{\partial_x}\right){\psi^{}_{aL}}\\
+ \frac{g}{N}\left({\psi^\dagger_{aR}}{\psi^{}_{aL}} +
{\psi^\dagger_{aL}}{\psi^{}_{aR}}\right)^2
\end{multline}
Here, $a=1,2,\ldots,N$. ${\psi^{}_{aR}}, {\psi^{}_{aL}}$
annihilate fermions of flavor $a$ at, respectively,
the right and left Fermi points.
This is the $U(N)$ Gross-Neveu model. Let us consider
its large-$N$ limit.

\begin{figure}[tbh!]
\vskip 1 cm
\includegraphics[width=3.25in]{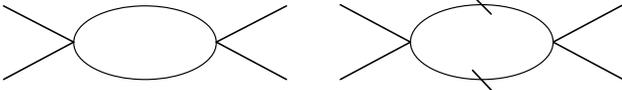}
\caption{One-loop diagrams contributing to the RG equation
for $g$.}
\label{fig:GN-one-loop}
\end{figure}

First, consider the RG equation for $g$. There is a
single one-loop diagram, depicted on the left of fig. \ref{fig:GN-one-loop}.
All other diagrams are smaller by powers of $1/N$.
The resulting RG equation is:
\begin{equation}
\label{eq:GN-rgeqn}
\frac{dg}{d\ell} = \frac{g^2}{2\pi}
\end{equation}

If we introduce a Hubbard-Stratonovich field $\Delta$
which decouples the interaction
\begin{multline}
{\cal L} = {\psi^\dagger_{aR}}i\left({\partial_t}+{\partial_x}\right)
{\psi^{}_{aR}}
+ {\psi^\dagger_{aL}}i\left({\partial_t}-{\partial_x}\right){\psi^{}_{aL}}\\
+ \Delta\left({\psi^\dagger_{aR}}{\psi^{}_{aL}} +
{\psi^\dagger_{aL}}{\psi^{}_{aR}}\right)
+\frac{N}{2g}\,{\Delta^2}
\end{multline}
then we can integrate out $\psi$:
\begin{equation}
\label{eq:Delta-action}
{\cal L}_{\rm eff} = 
\frac{N}{2g}\,{\Delta^2} -
N\,{\rm tr}\ln\left(\left(i{\partial_t}-i{\partial_x}{\sigma_3}
+\Delta{\sigma_1}\right)\delta_{ab}\right)
\end{equation}
where we have combined ${\psi^{}_{aR}}, {\psi^{}_{aL}}$
into a two-component spinor on which ${\sigma_{1,3}}$
act. $\Delta$ cuts off the infrared logarithmic
divergence of the ${\rm tr}\ln$ term.
\begin{eqnarray}
\label{eq:Delta-action2}
{\cal L}_{\rm eff} &=& 
\frac{N}{2g}\,{\Delta^2} -
\frac{1}{4\pi} N\,{\Delta^2}\,\left(\ln\left(\Lambda/\Delta\right)-1\right)\cr
&=& \left(\frac{N}{2g}-\frac{1}{4\pi}\ln\left(\Lambda/\Delta\right)
+\frac{1}{4\pi}\right){\Delta^2}
\end{eqnarray}
where $\Lambda$ is the ultraviolet cutoff.
The second line suggests that $\frac{1}{4\pi}\ln\left(\Lambda/\Delta\right)$
renormalizes $1/g$. Thus, the effective action
manifestly embodies the physics contained in the RG
equation (\ref{eq:GN-rgeqn}). This is also clear from the close kinship
between the one-loop diagram which gives the ${\rm tr}\ln(\ldots)$
in (\ref{eq:Delta-action}) and the one-loop diagram which yields
the RG equation (\ref{eq:GN-rgeqn}).

However, strictly speaking, the formal relationship between the RG
equation and the effective action (\ref{eq:Delta-action2})
follows from the cutoff-independence of physically-measurable
quantities, $\Delta$ being the simplest. From (\ref{eq:Delta-action2}),
we see that the saddle-point value of $\Delta$ is
\begin{equation}
\Delta = \Lambda\,e^{-2\pi/g}
\end{equation}
Differentiating both sides with respect to
$\Lambda$ and requiring that $d\Delta/d\Lambda$ vanish,
we find:
\begin{equation}
\Lambda\frac{dg}{d\Lambda} = - \frac{g^2}{2\pi}
\end{equation}
which is equivalent to the RG equation (\ref{eq:GN-rgeqn}).

Now, consider a slightly more complicated situation:
\begin{multline}
{\cal L} = {{\cal L}_0}(\psi,{\psi^\dagger})
+ {{\cal L}_0}(\chi,{\chi^\dagger}) + {{\cal L}_{\rm int}}(\chi,{\chi^\dagger})\\
+ \frac{g}{N}\left({\psi^\dagger_{aR}}{\psi^{}_{aL}} +
{\psi^\dagger_{aL}}{\psi^{}_{aR}}\right)^2\\
+ \frac{v}{\sqrt{N}} \left({\psi^\dagger_{aR}}{\psi^{}_{aL}} +
{\psi^\dagger_{aL}}{\psi^{}_{aR}}\right)
\left({\chi^\dagger_{aR}}{\chi^{}_{aL}} +
{\chi^\dagger_{aL}}{\chi^{}_{aR}}\right)
\end{multline}
where ${{\cal L}_0}(\psi,{\psi^\dagger})$ is
the same as the first line of (\ref{eq:GN-model}).

The RG equation for $g$ now reads:
\begin{equation}
\label{eq:GN'-rgeqn}
\frac{dg}{d\ell} = \frac{g^2}{2\pi} + \frac{v^2}{2\pi}
\end{equation}
The second term arises from the diagram
on the right-hand-side of fig. \ref{fig:GN-one-loop}.

\begin{figure}[tbh!]
\vskip 1 cm
\includegraphics[width=3.25in]{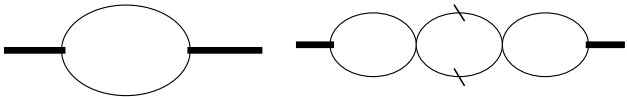}
\caption{The one- and three-loop contributions to the effective
action for $\Delta$ which encapsulate the same
physics as the one-loop RG equation.}
\label{fig:1-3-loop-toy}
\end{figure}

Now, consider ${\cal L}_{\rm eff}(\Delta,\chi,{\chi^\dagger})$,
where $\Delta$ is introduced to decouple the
$g$ interaction, as before.
We can no longer calculate it exactly. However,
if we assume that $v\ll g$,
we can calculate it to lowest non-trivial
order in $v$. The lowest order diagram by which it
contributes is the second one of fig. \ref{fig:1-3-loop-toy}.
We integrate out $\psi$ completely
and integrate out all $\chi$ fluctuations
above some cutoff $\Lambda'$. We are free
to choose $\Lambda'$ as we like, of course,
but the natural choice is $\Lambda'=\Delta$
so that our effective action is for excitations
with energies of order $\Delta$.
The resulting effective action is
\begin{multline}
\label{eq:Delta-action3}
{\cal L}_{\rm eff} = 
\frac{N}{2g}\,{\Delta^2} -
\frac{1}{4\pi}N\,{\Delta^2}\,\ln\left(\Lambda/\Delta\right)\\
-\: \frac{1}{4\pi}{v^2} {N} {\Delta^2}
\left(\ln\left(\Lambda/\Delta\right)\right)^3\\
+\:{{\cal L}_0}(\chi,{\chi^\dagger}) + {{\cal L}_{\rm int}}(\chi,{\chi^\dagger})
\end{multline}
The last term looks strange. It looks like an $O\left({v^2}\right)$
renormalization of $1/g$, which would imply a contribution of
order ${g^2}{v^2}$ to $dg/d\ell$. Also, this term is proportional
to $\left({\Delta^2}\ln\left(\Lambda/\Delta\right)\right)^3$ rather than
${\Delta^2}\ln\left(\Lambda/\Delta\right)$ as we might have expected from
the second diagram of fig. \ref{fig:GN-one-loop} and the RG equation
(\ref{eq:GN'-rgeqn}).

The problem is that we are looking at ${\cal L}_{\rm eff}(\Delta)$,
which is not observable, rather than the vacuum expectation value
of $\Delta$, which is. When we compute $\Delta$, the two
difficulties noted above resolve each other. Minimizing
the effective action, we find that, to lowest non-trivial
order in v:
\begin{equation}
\ln\left(\Lambda/\Delta\right) = \frac{2\pi}{g} - \frac{2\pi\,v^2}{g^3}
\end{equation}
or
\begin{equation}
\Delta = \Lambda \, e^{-2\pi/(g + {v^2}/g)}
\end{equation}
Requiring $\Lambda$-independence of $\Delta$, we recover
the RG equation for $g$ -- and also the RG equation for
$v$ which we didn't compute.

The lesson is that a one-loop contribution to the
RG equation for a coupling constant might manifest
itself in an effective action only through a higher-loop
(in the above example, three-loop) diagram. In the
interacting electron problem in a random potential,
diagrams (a), (b), and (c) could enter $S_{\rm eff}[X]$
through three-loop diagrams.

%\bibliography{../../bibs/corr}

\begin{thebibliography}{99}
\bibitem{resistance-note} Note that Finkelstein uses
$g$ to denote resistance, as did Chamon, Ludwig, and Nayak,
in keeping with his notation. However, the more widespread
usage is for $g$ to denote conductance. To avoid confusion,
we adopt the notation of Castellani, Di Castro, Lee, and Ma:
$t$ denotes resistance.
\bibitem{sdw-note} When the bare interaction is weak,
disorder does not allow the former, an SDW, and
its necessary concomitant, a CDW,
to form since the energy scale at which the
interaction becomes strong is small compared to
the inverse of the elastic scattering time.
On the other hand, one can imagine that
strong interactions could lead to the
formation of an SDW in the ballistic regime,
i.e. at length scales shorter than the mean-free path.
At longer scales, SDW order would be
disrupted by impurities, but the correlation length could
be long if the impurity density is low.
In the problem which we consider, however,
interactions are assumed to be weak in the ballistic regime.
They only become strong in the diffusive regime, where
SDW order will not develop.
\end{thebibliography}
%\bibliographystyle{prsty}

\end{document}